# The Audible Artefact: Promoting Cultural Exploration and Engagement with Audio Augmented Reality

Laurence Cliffe[1], James Mansell[2], Joanne Cormac[3], Chris Greenhalgh[1] and Adrian Hazzard[1]
[1]School of Computer Science, [2]Department of Cultural, Media and Visual Studies, [3]Department of Music
The University of Nottingham, United Kingdom
laurence.cliffe, james.mansell, joanne.cormac, christopher.greenhalgh, adrian.hazzard@nottingham.ac.uk

## ABSTRACT

This paper introduces two ongoing projects where audio augmented reality is implemented as a means of engaging museum and gallery visitors with audio archive material and associated objects, artworks and artefacts. It outlines some of the issues surrounding the presentation and engagement with sound based material within the context of the cultural institution, discusses some previous and related work on approaches to the cultural application of audio augmented reality, and describes the research approach and methodology currently engaged with in developing an increased understanding in this area. Additionally, it discusses the project within the context of related cultural and sound studies literature, presents some initial conclusions as a result of a practice-based approach, and outlines the next steps for the project.

## 1 INTRODUCTION

In this paper we present two ongoing projects, *Mapping the Symphony* and *National Science and Media Museum Gallery Listening Sessions* both of which attempt to directly apply audio augmented reality as a means of promoting visitor exploration and engagement with art, artefacts, sound archive material and their related stories.

Within the context of this paper, audio augmented reality (AAR) is considered as a virtual audio augmentation of the physical and visual reality, or the physical artefact. In an approach similar to [1, 2, 3], a virtual audio soundscape currently replaces the ambient acoustic reality of the location, rather than mixing with it, a mixed reality experience is therefore realised through the meeting of physical artefact and virtual audio. The possible inclusion of external acoustic ambience as a part of the system is discussed in section 6.3.

We show how this practice-based, research through design approach has developed a workable object detection and nomadic indoor positioning prototype that extends the capabilities of art and artefacts to advertise their presence to visitors through audio augmentation.

We also outline a methodological approach, where the ethnographical study and subsequent ethnomethodological analysis [4, 5] of deployed prototypes will support the developing iterations of the project going forward [6].

Additionally, we discuss the project's potential for increasing public engagement with sound archive material, art and artefacts in relation to the system's ability to extend the communicative potential of the museum and gallery object, and in relation to affording primacy to the sonic, rather than the visual. This is based on related contemporary sound studies literature, historical examples of augmented reality intervention within cultural institutional contexts, and related work.

Also described is how the project builds on some of the approaches outlined by Zimmerman & Lorenz in relation to the LISTEN system [2], namely the concept of the *attractor sound*, to develop an approach capable of longer range indoor positioning with a reliance on virtually no background technological infrastructure.

Finally, we outline the ongoing, upcoming and future work related to the project, and present a sound art installation environment for each of the projects that will act as settings within which we can conduct our ethnomethodological studies.

## 2   INTRODUCING THE PROJECTS

In the academic literature on museums, sound has been identified as having the potential to give exhibitions emotional power [7] and to generate multiplicity of interpretative perspective [8, 9]. The argument, in short, is that sonic exhibitions might help us to break from the truth effects of visual and textual storytelling and all of the asymmetrical power relations that they have been said to produce (especially in Foucauldian critiques of museums), opening the ground for visitors to 'poach' what they need from exhibitions, to borrow Boon's paraphrasing [8] of Michel de Certeau. Museums have enthusiastically embraced the challenge of sound, identifying its potential to produce more entertaining exhibitions, most notably in order to deal with auditory subject matter as in the case of the V&A's exhibitions 'David Bowie Is' and 'Pink Floyd: Their Mortal Remains' both of which provided a fully sound-tracked experience on headphones. Also of note is the Wellcome Collection's less obviously crowd-pleasing 2016 exhibition 'This is a Voice' which used installed sound, mainly via contemporary art commissions, to tell the scientific, medical and cultural story of the human voice.

This trajectory has established sound as an interpretation tactic in museums. However, there remains a live question about how to approach sound itself as an object of display. Kannenberg [10] has prompted us to think of sound as artefactual and worthy of display in museums (his own Museum of Portable Sound, a collection of everyday field recordings stored on an iPhone, is an effort in this direction). An exhibition at a major UK national museum *about* sound in one form or another, with visual and textual interpretation there as support rather than the main attraction, has yet to be achieved. Something of this nature was proposed by Boon et al [11] in preparation for a Science Museum exhibition on music. The proposed exhibition would seek to display the museum's collection of historical acoustic and sound reproduction technologies by foregrounding the listening experience and skills of those who designed and used objects such as tuning forks, gramophone sets and noise meters in the past. It would do so by engaging visitors primarily as listeners, drawing them into the auditory world of past music and sound technologies and leading them through the exhibition via the ear rather than the eye. That an exhibition such as the one described by Boon et al [11] has yet to be realised speaks, it is fair to say, to the enormous practical challenges of delivering this format.

This paper introduces two ongoing projects that aim to uncover the potential challenges and opportunities involved in the implementation of AAR as a means of promoting visitor exploration and engagement with cultural institutions, collections and exhibitions. Both projects also intend to explore how AAR can be deployed within cultural institutions, cultural venues and at heritage sites, and what value it may hold as a curatorial and artistic tool within these contexts.

### 2.1   National Science Museum Listening Sessions

At Science Museum, like many other science and technology museums, technological artefacts are conserved to maintain physical integrity but not to continue in an operational state. The acoustic world produced by late Victorian phonography, for example, is lost in national museum collections because to operate an original wax cylinder would risk its permanent degradation as a museum object. The situation is worse still for objects from the early electrical age: practices of conservation typically do not address the preservation of electronic function, meaning that a 1930s radio or gramophone set in the Science Museum collection cannot be switched back on. Paradoxically, Science Museum originally collected many of these objects, as Rich [12] has shown, specifically for the purpose of auditory demonstration. There is growing recognition that the history of sound recording and reproduction is a story worth telling in the museum. The challenge of doing so, with accessioned objects which cannot be used to recreate the sound worlds of the past, is still to be overcome.

There is also the challenge posed by the curious practice of collecting media technology and media content separately. The national sound archive is now held at the British Library, at one remove from the objects which once created and replayed recorded sound held largely at Science Museum and its regional branches, especially the National Science and Media Museum in Bradford. In response to the rapidly deteriorating physical state of British Library sound archive materials and others like it in regional collections, the Library has embarked on an ambitious programme of digitisation known as *'Unlocking Our Sound Heritage,'* though there remains little sense of what public use will be made of this digital archive once it is made available. From a silenced collection of sound technology hardware to an abundant, even noisy, digital sound archive, there is at present little strategy or consensus about what might be termed 'sonic engagement' – the practice of engaging the public in the history of hearing, listening and sound. The question of what sonic engagement should mean and how it should be achieved in the context of museums of science and technology was taken up by the Gallery Listening Sessions project at the National Science and Media Museum.

### 2.2   Mapping The Symphony

The second project introduced here explores how to present the evolving, fluid nature of music, exploding typical stereotypes surrounding classical music, and particularly the genre of the symphony, as something that is fixed and finished and must be performed as according to the composer's wishes. Museums have attempted to tell various stories about music history. In 2018, the V&A's *Opera: Passion, Power and Politics* exhibition traced the development of opera from the 17[th] century until today across Europe and America. This immersive experience brought performance objects, such as scenery and costumes to life through music. However, a different challenge is to explain how one piece of music can evolve due to changes made by editors, performers, arrangers, and the influence of outside forces, such as the resources available or the fashions prevalent in a particular time or place. It intends to explore how technology can guide the listener along the geographical and evolutionary journey made by a single symphony, using the musical content to engage visitors in its history, for example how they evolved to be performed as hymn settings in church, within ballets and operas in the theatre, and as military and brass band tunes in promenade and open-air concerts.

## 3　RELATED WORK

In addition to the exhibition based audio experiences outlined within the introduction of this paper, there are a number of other related projects that provide useful reference points.

Zimmerman & Lorenz's LISTEN system [2] provides an excellent example of the capabilities of AAR within the context of a cultural institution. The LISTEN project, which they describe as *'an attempt to make use of the inherent everyday integration of aural and visual perception'*, delivers a personalised and interactive location based audio experience based on an adaptive system model. It does this by tracking aspects of the visitors behaviour (which artworks have been visited, how long were they visited for etc.) to assign the visitor a behavioural model and adjust the delivery of audio content accordingly. The LISTEN system relies on a substantial technical background infrastructure to realise this personalised and invisible technical frontend experience for the visitor, who can wander freely through the exhibition space with just a set of customised headphones. LISTEN also introduces the concept of the *attractor sound*, which, based on the visitor's personalised profile model, suggests other nearby artworks to the visitor that may be of interest to them via spatially located audio prompts. Furthermore, LISTEN characterises many of the key differences between the usual audio guide experience and an interactive, adaptive and immersive approach. These include the dynamic delivery of spatial audio based on the listener's movement, and the delivery of related audio content based on the listener's proximity to an exhibit.

Like *The Rough Mile* [13], Sikora et al's archaeological AAR experience [3] could be categorised as an example of transformative soundscaping, where virtual audio is used to alter, or to reframe, rather than to directly compliment, the context of the locative experience. In the case of Sikora et al's AAR experience this change of context is from rural to urban. Being an outdoor experience it relies, as does *The Rough Mile*, on GPS technology for determining the position of the user within the physical landscape, values from which are translated into coordinates on a virtually authored representation of this landscape based on satellite imagery, onto which are placed virtual sound sources for the user to encounter in the real world. A similar authoring approach is taken by the system presented here, though being for indoor experiences they rely on a custom indoor positioning approach rather than GPS for determining the position of the user in virtual and physical space.

Seidenari et al's work on an automatic context-aware audio museum guide [14] demonstrates how a combination of both context modelling and artwork detection work together to influence the playback of audio descriptions. It also shows how the current object of the visitors focus is determined by a wearable camera based object recognition system. Additionally, the inclusion of speech detection within Seidenari et al's context-aware audio guide suggests a desire for users of such systems to maintain the ability to socially interact with their co-visitors, or rather it tries to ensure that *visitors can still talk to other visitors.* This ability is maintained in addition to an understanding that personalisation is a key factor in enabling museums to *talk with* visitors, rather than *talking to* them.

## 4　APPROACH AND METHODOLOGY

Both projects employ a practice-based, research through design approach where a series of iterative prototype interactive sound installations are developed through a cyclical process of development, deployment, study, analysis and redevelopment [6]. Ethnographical studies and subsequent ethnomethodological analysis [4, 5] of the deployed prototypes, where experts and prospective audiences are invited to participate and interact with the technologies within the installation environment will be undertaken. These study and analysis phases will then play a key role informing the subsequent redevelopment phases. The experts and prospective audiences actions and interactions will be observed, recorded and analysed in accordance with recognised ethnomethodological techniques, including the development of *thick descriptions* and a detailed understanding of *the machinery of interaction* [4, 5]. Additional data in relation to the audience experiences will be obtained from post-participatory questionnaires and interviews, along with quantitative analysis of data obtained from system logs and questionnaires.

## 5　SYSTEM DESCRIPTION

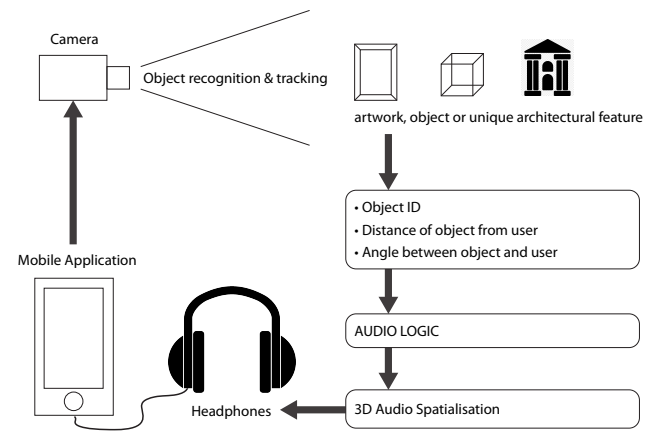

**Figure 1: Architecture of the current prototype**

The current prototype installations are delivered to listeners though headphones connected to a smartphone running an application which is authored using Unity [15], FMOD [16] and the Vuforia SDK [17], see figure 1.

Each sound source either has an audio logic script attached to it, or is attached to an FMOD event, which is provided with the current distance and orientation values of the listener in relation to it, which it uses to control the delivery of the audio source to the listener. This includes its spatial position within the virtual soundscape, based on the listener's orientation in relation to the virtual sound source and the real-world object, and its attenuation within the virtual soundscape, based on the listener's distance from the virtual sound source and the real-world object. The spatial position and the attenuation of the sound source within the stereo binaural mix of the virtual soundscape are the primary audio logic parameters which all the sound sources contain in order to place them within, and construct, a convincing and viable interactive and virtual three-dimensional soundscape. Based on these orientation and distance values other audio logic events can be scripted, such as the delivery of different audio files, or sections of an audio file, based on the listener's position in relation to the source.

In an attempt to develop a useable Bluetooth beacon based Indoor Positioning System (IPS), similar to that presented by Gimenes et al. [18], an initial prototype utilised a Puck.js [19], a Javascript programmable Bluetooth Low Energy (BLE) [20] beacon device, which was mounted on top of a set of Bluetooth enabled headphones. It was envisaged that this headphone-mounted interface would be useful in determining the user's real-world orientation in order to deliver an interactive surround sound experience. The Puck.js is equipped with a magnetometer, which was calibrated to work as a compass and return the user's bearing over a Bluetooth web connection to a local laptop running a Web Audio API [21] based web application. Bluetooth beacons were placed above wall mounted artwork, and a beacon scanning script was uploaded to the Puck.js which returned the RSSI values of nearby beacons to the local web application in an attempt to determine the listener's proximity to the artwork. Preliminary lab based testing and observations determined that this initial attempt was prone to ambient magnetic interference with the bearing data, along with problems of battery longevity in the Puck.js when using a frequency that gave a useable measurement of distance when returning the RSSI values back to the web application.

Informed and inspired by the artwork detection project presented by Seidenari et al. [14], and a realisation of the need to employ image recognition technology as a means to develop an application that was useable from both an authoring and curatorial perspective in a variety of locations, the Vuforia SDK [17] was adopted as a means to realise this. Along with artwork recognition, the use of image recognition and tracking technology also presented opportunities for the development of an Indoor Positioning System (IPS).

The Vuforia SDK enables the development of mobile augmented reality applications that use computer vision technology to recognise and track image targets and three-dimensional objects in real-time, and is compatible with both the iOS and Android mobile application platforms.

The Vuforia Engine's camera-based object recognition and tracking capabilities not only facilitate the recognition of the artwork and artefacts to which virtual audio sources can be associated, but also enable the implementation of an IPS were the mobile listener's angle and distance can be determined in relation to tracked, stationary two or three-dimensional objects.

Through an authoring approach similar to the one presented in the LISTEN system by Zimmerman & Lorenz [2], where a *world model* is combined with a *locative model*, we can determine our listener's position both in the physical and virtual environment of the experience. Additionally, the system is also capable of determining the listener's current focus by returning the angle and distance of the listener in relation to the tracked object.

An additional and important feature of this camera based IPS is made possible through Vuforia's *Extended Tracking* or *Simultaneous Localisation and Mapping* (SLAM) capability, delivered through either Apple's ARKit [22] or Google's ARCore [23], when compiled for delivery as either an iOS or Android application respectively. Vuforia's extended tracking enables the continued recognition and estimated location of a tracked object outside of the camera's field-of-view. This fusion based sensing technology extends our ability to determine the location of our physical objects and their associated virtual audio sources in relation to the listener's position in space. By being able to estimate both the angle and distance of the virtual audio sources around the listener, we can deliver a virtual and interactive three-dimensional soundscape based on the listener's physical, real-world environment.

Initial prototype designs centred around tracking the objects to which the virtual sound sources where going to be attached to, and using these as reference points to determine our listener's position and orientation, an approach that seemed natural given that these were the objects that we wanted to detect. But through the prototype development stages, once a system had been developed that demonstrated a useable degree of accuracy and reliability, and through the trials and manipulations involved in sculpting the positions and dimensions of the virtual audio sources in physical space, a 'natural feature' detection approach emerged. This approach involved providing the object tracking software (Vuforia) with isolated images of unique and static physical features within the experience environment, and determining the listener's position and orientation in relation to these physical features, and in-turn determine the position of the user in relation to the object to be augmented with sound.

## 6 DISCUSSION

### 6.1 Beyond line-of-sight

The fact that sound has the ability to extend the communicative reach of the visual is a theme reflected upon by both Conner [24] and Attali [25]. In essence, we are made aware of objects and events that emit sound prior to observing them; hearing is a sense that augments our vision.

This camera-based, natural feature detection approach to indoor positioning is both different and interesting. By associating virtual and spatialised audio sources to objects, or features, that may not be directly related to the experience, one can begin to think about how artworks or artefacts within gallery and museum spaces could advertise their presence beyond the traditional confines of line-of-sight.

For example, detecting and tracking a painted portrait in the gallery lobby could advertise, through spatially positioned virtual audio, the presence of the natural history exhibits through the door to the left (bird song, monkeys and lions) and a contemporary urban photography exhibition through the door to the right (a city soundscape).

Additionally, one can see how this could also be used to curate and design visitor journeys through an exhibition or a collection of artefacts by triggering sound sources at certain times in certain locations, or in relation to other objects. Also possible would be advertising the location of other objects in relation to the one you are currently viewing, *associated objects* that work well together sonically as well was contextually, guiding and suggesting potential trajectories to the listener.

Such approaches build upon Zimmerman & Lorenz's [2] concept of the *attractor sound*, a feature of the LISTEN system that recommends additional artworks, via emitted and localized virtual sound sources, to users of the system based on adaptive and personalised recommendations.

Giving objects within these cultural spaces the ability to communicate to visitors beyond line-of-sight has the ability to provide great potential, and significant challenges, for the designers and curators of such spaces. Spaces where the visual has maintained primacy from architectural design through to curatorial decision making for centuries [24], and constitutes a new way of seeing within such environments.

Additionally, regarding the authorship and curation of trajectories through an exhibition space, such an approach places the object in the role of both waypoint and content, making an object potentially a *functional*, and a *thematic*, part of the system. This is considered in relation to Hazzard et al's definition of an audio event that *'tells, or supports the telling of the narrative'* as being thematic, and an audio event that *'supports participants in their navigation or comprehension of the experience'* as being a functional one [10].

### 6.2   The Permeable Institution

The idea of the permeable cultural institution is made with reference to the *We AR* group's subversive, site-specific and visual based augmented reality interventions at MoMA in 2010 [26]. This artist group curated an unofficial, virtual art exhibition and virtual artworks were placed within MoMA's gallery spaces which visitors could view through their smartphones, leading Thiel [26] to remark that *'The institutional walls of the white cube are no longer solid…'*.

In relation to the application of audio augmented reality that has already been discussed, we can see how this also applies through the potentially subversive or unofficial use of virtually *placed sound* [27] within an institution. But, in the case of AAR, Thiel's remarks take on a more literal meaning, both in relation to the internal walls of the gallery and its external boundaries, and present opportunity as well as challenge. If we can hear the visual before we see it, then the dividing walls of gallery rooms are no longer obstacles to our exploration, if we can hear the contents of the institution before we arrive, then these objects are no longer confined by external architecture. According to Breitsameter, in Behrendt [27], this fluid and borderless design approach stems from '*a sonic understanding of space*' which allows for a space which is more permeable and one that *'doesn't suggest the same kind of hard and fast boundaries of a visual construction of space'*.

In relation to the appropriation of experimental artistic interventions for institutional based curatorial purposes, it is perhaps worth noting Zimmerman & Lorenz's positive curatorial feedback on their LISTEN system [2] which acknowledges the curatorial potential of innovative, less descriptive and enriched audio content. Which also speaks more generally for the curatorial appropriation of experimental sonic art practice as a tool for cultural engagement within the gallery and museum.

### 6.3   Ambient Inclusion

The locative audio walk *The Rough Mile* [13] provides compelling evidence, and testimony from its participants, for the inclusion of situated ambient noise within such an experience, which, in this particular case, was realised through the use of bone-conducting headphones. Participants of *The Rough Mile* generally noted that their ability to hear both the real and virtual sound sources of the experience's location added to their feeling of immersion within the experience, and that it better situated the virtual sound within the physical location. Although there is also evidence to suggest that occasional loud ambient sounds masked the ability to hear the virtual audio through the bone-conducting headphones. This problem has also been observed whilst using bone-conducting headphones with this project's prototypes, along with a significant loss in the fidelity of the audio, which can make it difficult to discern nuances in the spatialisation and frequency of the audio signal, an issue not experienced when using more traditional closed-cup, or over-ear headphones. A logical 'best of both worlds' solution could be to include ambient sound within the experience either through the smartphone's microphone, or an external stereo or binaural microphone, delivered into the headphones in real-time, as a way of maintaining the immersive quality of a high fidelity surround sound experience, whilst adding to these immersive qualities though ambient inclusion. Additionally, this would enable the level of the two sources, the virtual recorded audio content and ambient audio content, to be monitored, mixed and managed.

## 7    NEXT STEPS

Participatory audience focus groups have been organised within the context of a gallery space where interactions between users, the space, the technology, and an audio augmented museum artefact can be observed and recorded. The first will take place at The National Science and Media Museum in Bradford where a 1940's radio receiver from the museum's collection will be augmented with archive BBC radio broadcasts from the same period, originally produced for a project called '*Sound and Fury: Listening to the Second World War*'. Participants will be able to 'tune in' to different broadcasts through their movement and proximity to the radio set within the gallery space (see figure 2).

The event will be documented, and an in-depth ethnomethodological analysis [13, 14] will be conducted in an effort to unfurl and build 'thick descriptions' of the social interactions around the participants involvement with the installation. It is also hoped that additional data can be gathered by conducting interviews with participants subsequent to their interacting with the experience, and also asking participants to complete a questionnaire regarding their experiences. The analysis of the results of these workshops will be considered throughout the development of the subsequent iteration of prototypes, in an effort to inform the ongoing design process within the context of a recognised, iterative practice-based research approach [12]. Future studies may also include a quantitative analysis of engagement, as demonstrated by Sikora et al. [3], where participants time spent at specific audio augmented locations using the AAR system is compared alongside time spent at these locations by participants not using the AAR system, with data being made available for analysis through observation and data logging within the AAR system.

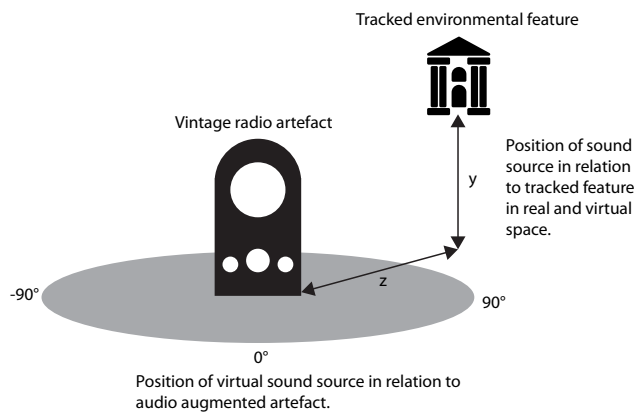

**Figure 2. Diagram of radio installation design for the Science Museum Listening Sessions.**

Similarly, an installation environment has been proposed for *Mapping the Symphony*. This will involve virtually attaching the musically rendered symphonic arrangements to 3D models of historical symphonic halls associated with those specific arrangements. With the 3D architectural models placed on their relevant geographical locations upon a map rendered on the gallery floor, visitors will be able to explore a virtual historical symphonic soundscape generated by their proximity to these models, from which will emanate the different symphonic arrangements, see figure 3.

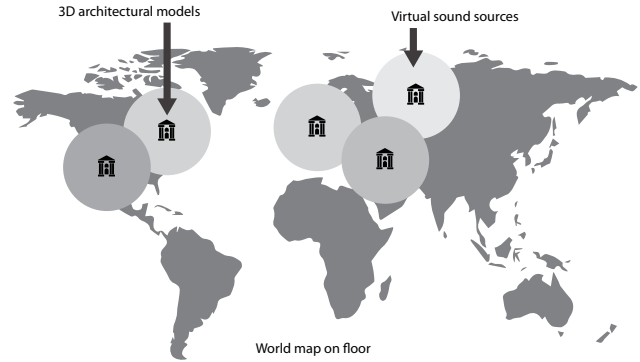

**Figure 3. Diagram showing an installation layout for *Mapping the Symphony* project.**

It is envisaged that both projects will be evaluating similar interactions in a similar manner, though within different locations and contexts, with a view to obtaining data that can help inform both the artistic and curatorial opportunities and challenges for the system across a variety of cultural and institutional contexts.

In relation to the IPS employed by the system, throughout the initial prototype development significant discrepancies have been observed in the accuracy of the tracking (both distance and orientation values) when the system switches to a reliance on SLAM based extended tracking, from tracking within the camera's field-of-vision. Various interpolations and algorithmic, value smoothing techniques have been used to improve this, though significant work remains in terms of developing these, and in terms of accounting for this within the design and authoring of the related experiences. Of particular interest would be how the reliability of tracked targets are affected by changes in lighting conditions and human traffic and how any shortcomings relating to this can be accounted for within the design and authoring process. Additionally, the inclusion of ambient noise as an additional sound source for these projects, whilst maintaining the quality of the interactive surround sound experience, offers much promise regarding increased immersion and adding to the listener's *suspension of disbelief* in the virtual soundscape.

## 8    CONCLUSIONS

An initial demonstration of the current prototype at Science Museum was received favourably by curatorial and collections staff in terms of its potential application within a museum context, and in terms of its immersive qualities. It is therefore believed that the approach to AAR outlined in this paper warrants further exploration, and demonstrates a potential for connecting silenced sound technology hardware in museum collections with relevant archive recordings, and in helping to engage visitors with these objects, their stories, and their associated audio archival material. It is also proposed that this approach demonstrates a contribution to indoor positioning within gallery and museum environments through the application of camera based object detection for determining visitor location and focus. Which, in turn, enables the system to expand upon Zimmermann & Lorenz's [2] concept of the *attractor sound* through a system reliant on little background infrastructure. Also concluded is that this appropriation and application of *Simultaneous Localisation and Mapping* (SLAM) for solely audio augmentation purposes works well with this particular technologies current shortcomings. The small and gradual movement of overlaid graphics placed on real world objects that can sometimes be observed with visual based AR applications are not so acute or obvious when translated to the spatial position of audio sources, a shortcoming which is the result of the mapping technologies adjusting their placement of virtual augmentations as they build up, or gain additional information about their environment [22, 23].

## ACKNOWLEDGMENTS

The author is supported by the Horizon Centre for Doctoral Training at the University of Nottingham (RCUK Grant No. EP/L015463/1) and Fusing Audio and Semantic Technologies (FAST).